\documentclass[pra,twocolumn]{revtex4-1}
\usepackage{bm, amssymb, physics, environ, tikz, float, amsmath, subcaption}
\usetikzlibrary{decorations.pathmorphing, shapes.misc, shapes.geometric,calc, arrows.meta, arrows, decorations.pathmorphing,fit}
\definecolor{Dark_Green}{rgb}{0,128,0}

\begin{document}
\title{Photon routing in disordered chiral waveguide QED ladders: Interplay between photonic localization and collective atomic effects}
\author{Nishan Amgain and Imran M. Mirza} 
\email{mirzaim@miamioh.edu}
\affiliation{Macklin Quantum Information Sciences,  Department of Physics,\\ Miami University, Oxford, Ohio 45056, USA}

\begin{abstract}
In recent years, photon routing has garnered considerable research activity due to its key applications in quantum networking and optical communications. This paper studies the single photon routing scheme in many-emitter disordered chiral waveguide quantum electrodynamics (wQED) ladders. The wQED ladder consists of two one-dimensional lossless waveguides simultaneously and chirally coupled with a chain of dipole-dipole interacting two-level quantum emitters (QEs). In particular, we analyze how a departure from the periodic placement of the QEs due to temperature-induced position disorder can impact the routing probability. This involves analyzing how the interplay between the collective atomic effects originating from the dipole-dipole interaction and disorder in the atomic location leading to single-photon localization can change the routing probabilities. As for some key results, we find that the routing probability exhibits a considerable improvement (more than $90\%$ value) for periodic and disordered wQED ladders when considering lattices consisting of twenty QEs. This robustness of collective effects against spontaneous emission loss and weak disorders is further confirmed by examining the routing efficiency and localization length for up to twenty QE chains. These results may find applications in quantum networking and distributed quantum computing under the realistic conditions of imperfect emitter trappings.
\end{abstract}
\maketitle

\section{\label{sec:I} Introduction}  
The burgeoning field of waveguide quantum electrodynamics (or, in short, wQED) provides exciting ways to study strong light-matter interaction between trapped quantum emitters and one dimension propagating photons \cite{sheremet2023waveguide, roy2017colloquium, ciccarello2024waveguide}. Thanks to the advancement in optical technologies, in the last decade or so using wQED architectures, many fascinating quantum effects have been experimentally observed and theoretically proposed. The list of effects is quite diverse and includes qubit-qubit entanglement \cite{gonzalez2014generation, mirza2016multiqubit,patrick2024chirality}, electromagnetically induced transparency \cite{witthaut2010photon, berndsen2023electromagnetically, long2018electromagnetically}, spin-momentum locking of light \cite{lodahl2017chiral, bliokh2015quantum, petersen2014chiral}, collective atomic emissions \cite{chang2018colloquium, asenjo2017exponential, goban2015superradiance}, photon-photon correlations \cite{fang2015waveguide, shen2015photonic, mahmoodian2018strongly}, and few-photon scattering \cite{hurst2018analytic, liao2016photon} (to name a few). On the experimental side, several platforms are now available to perform wQED experiments, for instance, superconducting qubits coupled with microwave transmission lines \cite{blais2021circuit}, nanowires interacting with quantum dots \cite{akimov2007generation}, and naturally occurring atoms (such as Ce atoms) coupled with nanophotonic waveguides \cite{turschmann2019coherent,liao2016photon}, etc. Another appealing feature of such architectures is their utilization in performing quantum-enabled information tasks, particularly in the context of long-distance quantum communications and networking \cite{kimble2008quantum, gisin2007quantum}.

Among the fascinating effects observed in wQED, spin-momentum locking is a relatively recent development that has opened up a new research direction of chiral quantum optics \cite{lodahl2017chiral}. Since its experimental demonstration, chiral light-matter interfaces have shown to play a vital role in enhancing entanglement \cite{mirza2016two, gonzalez2015chiral, patrick2024chirality}, controlling the dispersion properties of photons in one- \& two-dimensional wQED \cite{mirza2018influence, mahmoodian2020dynamics, marques2021two}, and constructing uni-directional photonic devices \cite{guimond2020unidirectional}. Within this list, the application of chiral wQED in enhancing the routing efficiency of single photons in multi-port optical devices stands out in the context of applications in quantum networks. In this regard, Gonzalez et al. have shown that chirality can achieve deterministic routing for single photons when single three-level atoms are side-coupled to two waveguides in a ladder geometry \cite{gonzalez2016nonreciprocal}. Moving one step further, we have shown that the routing efficiency undergoes considerable reduction when the spontaneous emission from the two-level QE is considered. To address this issue, we proposed using collective atomic effects by introducing dipole-dipole interaction among QEs \cite{poudyal2020collective}. We found that the dipole-dipole interaction (DDI) provides a non-waveguide channel to establish strong emitter-emitter coupling, which protects the photon routing against environmental decoherence.

\begin{figure*}
\centering
\includegraphics[width=5.3in, height=2.35in]{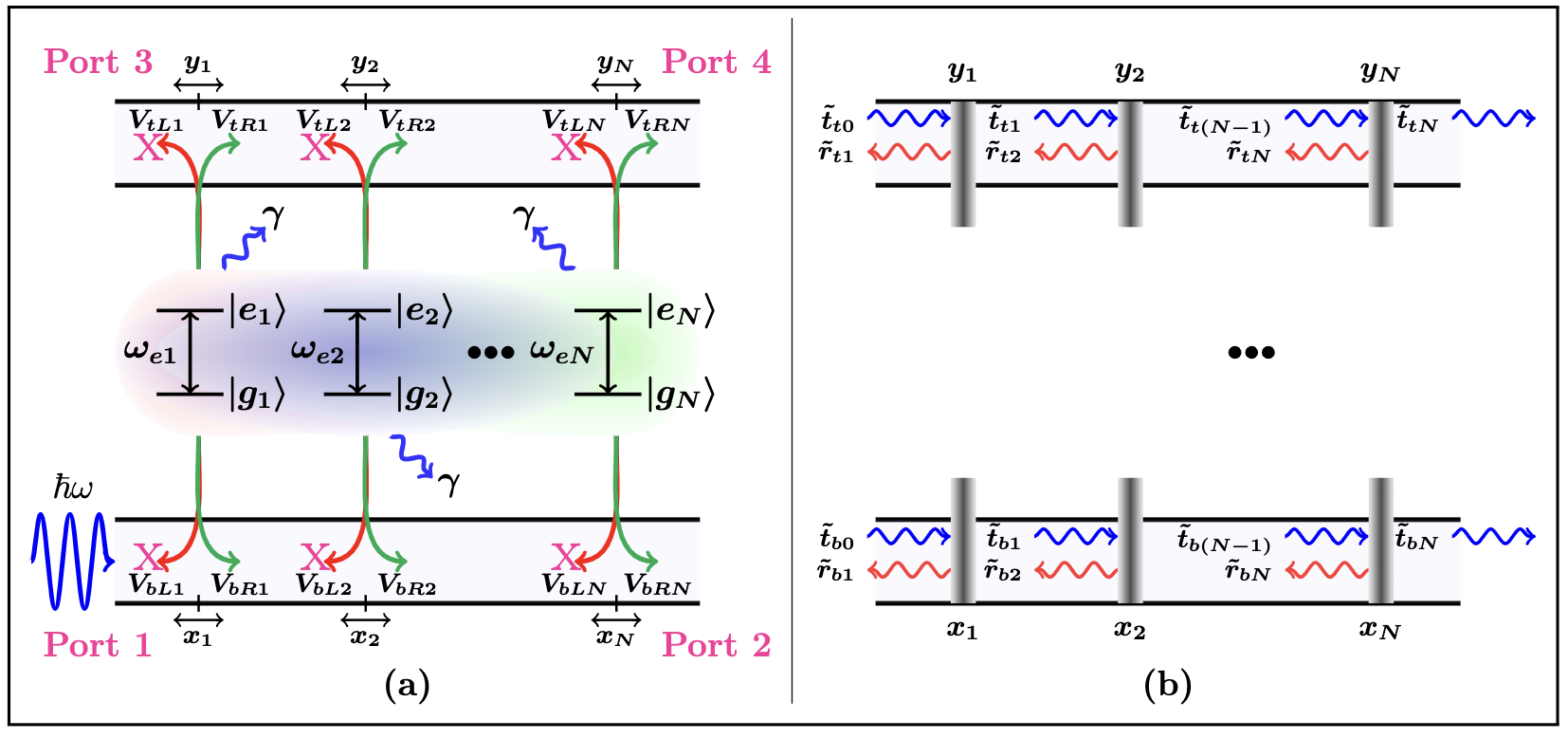} 
\captionsetup{
format=plain,
margin=1em,
justification=raggedright,
singlelinecheck=false
}
\caption{(Color online) (a) Illustration of the waveguide QED ladder model studied in this paper. A single photon with energy $\hbar\omega$ is launched from the left end of the bottom waveguide (Port 1), and the goal is to achieve the deterministic routing of the photon from Port 1 to Port 4. The blurry color spread across emitters shows the infinitely long-ranged DDI among QEs. This work assumes perfect chiral waveguides in which the backward or left emission directions are prohibited (as indicated by red crosses in the figure). (b) A schematic diagram indicating the transmission and reflection coefficients at each boundary where each QE resides. The blurriness in the boundaries indicates the position disorder. }\label{Fig1}
\end{figure*}

In the present work, we concentrate on an important scenario in which, due to a finite temperature in the environment or imperfect trapping, the QEs' locations are deviated away from their mean periodic positions. Such a random disorder, when sufficiently large, is known to completely halt the motion of electrons in condensed matter systems (the so-called Anderson localization) \cite{anderson1958absence, lagendijk2009fifty}. However, in recent years, similar behaviors have been reported in transport problems involving electromagnetic \cite{segev2013anderson, figotin1997localization}, acoustic \cite{condat1987resonant}, and spin waves \cite{igarashi1987anderson} as well. In the atomic and molecular physics domain, Roati et al. have studied Anderson location with Bose-Einstein condensates \cite{roati2008anderson}, Aspect et al. have examined the problem of localization with ultracold atoms \cite{aspect2009anderson}, and Marcuzzi et al. reported the impact of such position disorder on the facilitation dynamics in Rydberg gases \cite{marcuzzi2017facilitation}. The question we try to answer here is how such a position-disorder-induced photon localization can compete against the collective DDI to alter the single photon routing in many-body chiral wQED architectures. 

As for some of the noteworthy findings, we conclude that the process of routing (quantified in terms of desired port detection probability) tends to support more than $90\%$ routing as we consider the collective effects of fifteen or a higher number of QEs in the wQED ladder. Furthermore, this routing efficiency shows robustness against spontaneous emission loss and weak disorder in the locations of QEs. The localization length shows a decreasing trend as the disorder is enhanced (which is contrary to the standard Anderson localization behavior \cite{anderson1958absence, lagendijk2009fifty}), indicating the dominating nature of collective effects, which tends to suppress the process of photon localization for wQED ladders with large number of (i.e., around twenty) QEs.

The rest of the paper is structured as follows. In Sec.~\ref{sec:level2}, we lay out the theoretical model studied in this work, which involves the Hamiltonian and the single-photon quantum state specification. Next, in Sec.~\ref{sec:level3}, we report the equations of motion obeyed by the probability amplitudes and express these equations in terms of transport coefficients. Following that, in Sec.~\ref{sec:level4}, we report the results of our study, including transport probabilities, routing efficiency, and localization length. Finally, in Sec.~\ref{sec:level5} we close by summarizing the main conclusions of our work.

\section{\label{sec:level2} Theoretical Description}
\subsection{Model Hamiltonian}
As depicted in Fig.~\ref{Fig1}, our theoretical model consists of an array of $\mathcal{N}$ number of QEs with a ground state $\ket{g_j}$, excited state $\ket{e_j}$, transition frequency $\omega_{eg_{j}}$, location $x_j$ with respect to the bottom waveguide, $y_j$ with respect to the top waveguide, and spontaneous emission rate $\gamma_j$, $\forall j=1,2,...,\mathcal{N}$. This emitter chain is simultaneously coupled with two infinitely long one-dimensional waveguides with top and bottom waveguide coupling rates given by $V_{t\alpha_j}$ and $V_{b\alpha_j}$, respectively, with $\alpha$ representing either left (L) or right (R) direction in the waveguide. For simplicity, we suppose all QEs to be identical, with the ground state energy being set as a reference. Therefore, the free Hamiltonian of the QEs can be expressed as follows
\begin{align}\label{HamQE}
    \hat{\mathcal{H}}_{QE}=\sum^N_{j=1} \hbar\widetilde{\omega}_{eg_{j}}\hat{\sigma}^\dagger_j\hat{\sigma}_j,
\end{align}
where $\widetilde{\omega}_{eg_{j}}:=\omega_{eg_{j}}-i\gamma_j$ with $\hat{\sigma}_j$ being the $j^{th}$-QE lowering operator following the standard Fermionic anticommutation relation $\lbrace \hat{\sigma}_i,\hat{\sigma}^\dagger_j\rbrace =\delta_{ij}$. The spontaneous emission rate $\gamma_j$ has been inserted by hand here using the detailed analysis reported in Ref.~\cite{shen2009theoryI, shen2009theoryII}. The two waveguides in our work are modeled as quantum harmonic oscillators with infinitely many modes, which allows us to write their Hamiltonian in the real-space formalism of quantum optics \cite{shen2009theoryI, shen2009theoryII} as
\begin{align}\label{HamWav}
    \hat{\mathcal{H}}_W &= -iv_{bg}\hbar \int \left[\hat{c}_{bR}^{\dag}(x)\partial_x \hat{c}_{bR}(x)  -  \hat{c}_{bL}^{\dag}(x)\partial_x \hat{c}_{bL}(x)\right] dx\nonumber\\
    &-iv_{tg}\hbar \int \left[\hat{c}_{tR}^{\dag}(y)\partial_y \hat{c}_{tR}(y)  -  \hat{c}_{tL}^{\dag}(y)\partial_y \hat{c}_{tL}(y)\right] dy.
\end{align}
Here $\hat{c}_{t\alpha}(y)$ and $\hat{c}_{b\alpha}(x)$ represents the photon annihilation operator in the top and bottom waveguide at location $y$ and $x$ in the $\alpha$ direction, respectively. These operators follow the typical Bosonic commutation relations: $[\hat{c}_{t\alpha}(y), \hat{c}_{t\beta}(y')] = \delta_{\alpha\beta}\delta(y-y')$ and $[\hat{c}_{b\alpha}(x), \hat{c}_{b\beta}(x')] = \delta_{\alpha\beta}\delta(x-x')$. We have selected two different group velocities $v_{tg}$ and $v_{bg}$ for the top and bottom waveguide, respectively. 

Next, two types of interactions are possible in our model. One is the interaction between the QEs and the waveguide fields, while the second is the direct interaction among the QEs due to DDI. The Hamiltonian for the former type $\hat{\mathcal{H}}_{I}$ under the rotating wave approximation can be expressed as
\begin{align}\label{HamInt1}
&\hat{\mathcal{H}}_{I}=\hbar\sum^{N}_{j=1}\sum_{\alpha=L,R}\Bigg[\int dx\delta(x-x_j)\big(V_{b\alpha_j}\hat{c}^{\dagger}_{b\alpha}(x)\hat{\sigma}_j + H.c. \big)\nonumber\\
&+\int dy \delta(y-y_j)\big(V_{t\alpha_j}\hat{c}^{\dagger}_{t\alpha}(y)\hat{\sigma}_j+H.c. \big)\Bigg].
\end{align}
Here, the Dirac delta function specifies the location of the $j$th QE in the chain, and the abbreviation $H.c.$ stands for Hermitian conjugate. Finally, by assuming the inter-emitter separation to be smaller than the resonant field wavelength, we introduce an infinitely long-ranged and direct DDI among the emitters \cite{agarwal2012quantum}. The Hamiltonian for such a DDI is given by
\begin{align}\label{HamInt2}
&\hat{\mathcal{H}}_{\rm DDI}=\hbar\sum^{N}_{i=1}\sum\limits_{\substack{j=1 \\ \hspace{-6mm}j > i}}^N J_{ij}\big(\hat{\sigma}^{\dagger}_{i}\hat{\sigma}_j + H.c. \big).
\end{align}
The DDI parameter (characterizing the strength of the interaction) is known to be sensitively dependent on the inter-emitter separation, as we discuss in the following subsection.
\subsection{DDI parameter}
For inter-emitter separation $\mathcal{R}_{ij}$ the DDI parameter $J_{ij} \equiv J\left(\mathcal{R}_{ij}\right)$ is known to obey the following form \cite{cheng2017waveguide}
\begin{align}\label{DDI}
J_{ij} &=\frac{3\Gamma_0}{4}\bigg(\frac{\cos \mathcal{R}_{ij}}{\mathcal{R}^3_{ij}}+\frac{\sin \mathcal{R}_{ij}}{\mathcal{R}^2_{ij}}-\frac{\cos \mathcal{R}_{ij}}{\mathcal{R}_{ij}}\bigg)\nonumber\\
&+\cos^2\theta\bigg(\frac{\cos \mathcal{R}_{ij}}{\mathcal{R}_{ij}}-\frac{3\cos \mathcal{R}_{ij}}{\mathcal{R}^3_{ij}}-\frac{3\sin \mathcal{R}_{ij}}{\mathcal{R}^2_{ij}} \bigg),
\end{align}
where the inter-emitter separation is defined as $\mathcal{R}_{ij}=\omega_{eg}|\vec{r}_i-\vec{r}_j|/c$, with $c$ being the speed of light, $\vec{r}_{i/j}$ identifies the position of $i/j$th emitter. In Eq.~\eqref{DDI}, $\Gamma_0$ represents the free space emitter decay rate. $\theta$ is the angle between the dipole moment vector $\vec{p}$ and the emitters relative position vector which is defined as $\cos\theta=\vec{p}\cdot(\vec{r}_i-\vec{r}_j)/\lbrace|\vec{p}||\vec{r}_i-\vec{r}_j|\rbrace$. Onward, we set $\theta=\pi/2$ for simplicity.
\begin{figure}
\centering
\includegraphics[width=3in, height=1.85in]{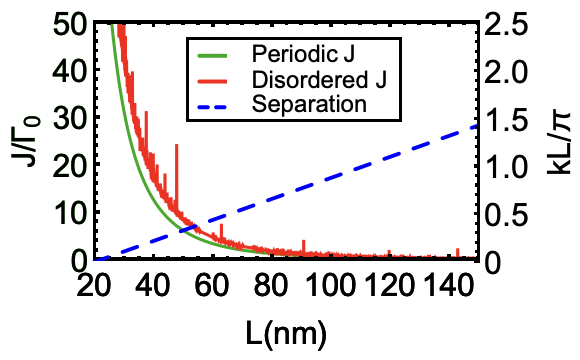} 
\captionsetup{
format=plain,
margin=1em,
justification=raggedright,
singlelinecheck=false
}
\caption{(Color online) The variation of DDI strength $J$ (in units of free space emitter emission rate $\Gamma_0$) as a function of the inter-emitter separation $L$ (measured in nanometers or nm). The results are obtained using Eq.~(\ref{DDI}). Green solid, red solid, and blue dashed curves represent the $J$ curve for the periodic case, $J$ curve for the disordered case, and variation of dimensionless separation $kL/\pi$. $k$ here represents the wavenumber corresponding to the emitter wavelength. For a quantum dot system, this wavelength has been reported to be $655{\rm nm}$ \cite{cheng2017waveguide}. We use a normal distribution for the disordered case with the mean of $\mu=L$ and the standard deviation (an indicator of the position disorder) equaling $0.2\mu$. The disordered curve is generated by averaging over $1000$ iterations of the numerical routine.}\label{Fig2}
\end{figure}
In Fig.~\ref{Fig2}, we plot the DDI parameter $J$ as a function of inter-emitter separation $L$ between two QEs. As expected from Eq.~(\ref{DDI}), despite being an infinitely long-range interaction, DDI decays quickly as the separation increases. To this end, we observe in periodic and disordered cases that as we approach $L=120nm$, we find $J\rightarrow 0$. Upon comparing the periodic case versus the disordered configuration, we find that for smaller to intermediate inter-emitter separation (i.e., $30nm\lesssim L\lesssim  60nm$), the DDI is slightly higher in the disordered case. For the disorder strength chosen here ($\sigma=0.2\mu$), we found that in the set of iterations in which the separation between the QEs becomes smaller than $\mu$, the DDI enhances considerably (as observed in the form of spikes in $J$ curve for certain separations in Fig.~\ref{Fig2}). Since the impact of such instances on the overall $J$ value was considerably higher than the iterations in which the random separation was larger than $\mu$, overall DDI remains higher. Note that for larger disorders (i.e., when $\sigma >0.2\mu$), the variation in $J$ fluctuated immensely. Therefore, we identify $\sigma=0.2\mu$ as the border value between the strong disorder case (when $\sigma >0.2\mu$) and the weak disorder case (when $\sigma<0.2\mu)$. 
\subsection{The single-excitation quantum state}
The quantum state of our system restricted to the single-excitation sector of the Hilbert space can be expressed as
\begin{align}\label{eq:state}
    \ket{\Psi} &= \Bigg[\sum^{N}_{j=1} \mathcal{A}_j\hat{\sigma}_j^\dag + \sum\limits_{\alpha=L,R}\Bigg\lbrace\int \varphi_{b\alpha}(x)\hat{c}_{b\alpha}^\dag(x)dx\nonumber\\
    & + \int\varphi_{t\alpha}(y)\hat{c}_{t\alpha}^\dag(y)dy\Bigg\rbrace \Bigg] \ket{\varnothing},
\end{align}
where $\ket{\varnothing} = \bigotimes_j\ket{g_j} \otimes \ket{0_{bR}} \otimes \ket{0_{bL}} \otimes \ket{0_{tR}} \otimes \ket{0_{tL}}$ represents the ground state in which all QEs are unexcited and no photons in the top and bottom waveguide in both left and right direction.
\section{\label{sec:level3} Amplitude equations and transport coefficients}
To calculate the single-photon transport properties, we insert total Hamiltonian $\hat{\mathcal{H}}=\hat{\mathcal{H}}_{QE}$+$\hat{\mathcal{H}}_W$ + $\hat{\mathcal{H}}_{I}$ + $\hat{\mathcal{H}}_{DDI}$ by combining Eq.~\eqref{HamQE}, \eqref{HamWav}, \eqref{HamInt1} and \eqref{HamInt2}; and the single-excitation quantum state from Eq.~\eqref{eq:state} into time-independent Schr\"odinger equation $\hat{\mathcal{H}}\ket{\Psi}=\hbar\omega\ket{\Psi}$. Here, $\hbar\omega$ represents the energy of the single photon launched into Port 1 of our wQED ladder setup (see Fig.~\ref{Fig1}). We thus arrive at the following set of coupled differential equations obeyed by the probability amplitudes in the current problem
\begin{subequations}\label{AmpEqsN}
    \begin{align}
    &\mathcal{A}_j\left(\omega - \widetilde{\omega}_{eg_j}\right) = \sum_{i>j}\mathcal{A}_i J_{ij}+\sum_{i<j}\mathcal{A}_i J_{ji} + \sum_{\alpha=L,R} \Big[V_{b\alpha j}\times\nonumber\\
    & \varphi_{b\alpha}(x_j)+ V_{t\alpha j}\varphi_{t\alpha}(y_j)\Big],   \\
    &-iv_{bg}\frac{\partial \varphi_{bR}(x)}{\partial x} + \sum_j \mathcal{A}_j V_{bRj} \delta(x-x_j) =  \omega \varphi_{bR} (x), \\
    &+iv_{bg}\frac{\partial\varphi_{bL}(x)}{\partial x} + \sum_j \mathcal{A}_j V_{bLj}\delta(x-x_j)  =  \omega \varphi_{bL} (x),\\
    &-iv_{tg}\frac{\partial \varphi_{tR}(y)}{\partial y} + \sum_j \mathcal{A}_j V_{tRj}\delta(y-y_j)  =  \omega \varphi_{tR} (y),\\
    &+iv_{tg}\frac{\partial\varphi_{tL}(y)}{\partial y} + \sum_j \mathcal{A}_j V_{tLj}\delta(y-y_j) =  \omega \varphi_{tL} (y).
    \end{align}
\end{subequations}
Proceeding further, we convert the above amplitude equations into transport equations by introducing the following ansatzes (also commonly referred to as Bethe ansatzes in literature \cite{dinc2020multidimensional}). To this end, following the schematic diagram shown in Fig.~\ref{Fig1}(b), we divide the whole atomic chain into smaller compartments. For instance, the $j$th compartment occupies boundaries at $x_j$ and $x_{j+1}$ concerning the bottom waveguide and boundaries at $y_j$ and $y_{j+1}$ concerning the top waveguide. This construction allows us to express the field wavefunctions (or waveguide amplitudes) in the following fashion
\begin{subequations}\label{BethAnstz}
    \begin{align}
        \varphi_{bR}(x) &= \tilde{t}_{bj}e^{ikx}\left[\Theta(x-x_j)-\Theta(x-x_{j+1})\right],\nonumber\\
        &~~\forall~~x_j\leq x\leq x_{j+1}~~ \text{with}~~\tilde{t}_{b0}=1,\\
        \varphi_{bL}(x) &= \tilde{r}_{bj+1}e^{-ikx}\left[\Theta(x-x_{j})-\Theta(x-x_{j+1})\right],\nonumber\\
        &~~\forall~~x_j\leq x\leq x_{j+1}~~ \text{with}~~\tilde{r}_{bN}=0,\\
        \varphi_{tR}(y) &= \tilde{t}_{tj}e^{imy}\left[\Theta(y-y_{j})-\Theta(y-y_{j+1})\right],\nonumber\\
        &~~\forall~~y_j\leq x\leq y_{j+1}~~ \text{with}~~\tilde{t}_{t0}=0,\\
        \varphi_{tL}(y) &= \tilde{r}_{tj+1}e^{-imy}\left[\Theta(y-y_{j})-\Theta(y-y_{j+1})\right],\nonumber\\
        &~~\forall~~y_j\leq x\leq y_{j+1}~~ \text{with}~~\tilde{r}_{tN}=0,
\end{align}
\end{subequations}
with, at the $j$th boundary, $\tilde{t}_{bj}$, $\tilde{r}_{bj+1}$, $\tilde{t}_{tj}$, and $\tilde{r}_{tj+1}$ being the transmission coefficient in the bottom waveguide, reflection coefficient in the bottom waveguide, transmission coefficient in the top waveguide, and reflection coefficient in the top waveguide, respectively. The parameters $k=\left(\omega-\omega_{eg}\right)/v_{bg}$ and $m = \left(\omega-\omega_{eg}\right)/v_{tg}$ appearing in the exponential represent the wavenumbers of the single photon field (detuned with respect to the emitter transition frequency) while propagating in the bottom and top waveguides, respectively. 

As the next step in finding a solution to Eq.~\eqref{AmpEqsN}, we regularize the waveguide amplitudes on each boundary point as
\begin{subequations}
    \begin{align}
       \varphi_{b\alpha}(x_j) = \lim_{\epsilon \rightarrow 0}\left[\frac{\varphi_{b\alpha}(x_j - \epsilon) + \varphi_{b\alpha}(x_j + \epsilon)}{2} \right],\\
       \varphi_{t\alpha}(y_j) = \lim_{\epsilon \rightarrow 0}\left[\frac{\varphi_{t\alpha}(y_j - \epsilon) + \varphi_{t\alpha}(y_j + \epsilon)}{2} \right].
    \end{align}
\end{subequations}
Integrating the relevant waveguide differential equations from $x_j - \epsilon$ to $x_j + \epsilon$ for the bottom waveguide and from $y_j - \epsilon$ to $y_j + \epsilon$ for the top waveguide with $\epsilon \ll 1$ yields the following set of transport equations 
\begin{subequations}
    \begin{align}
        &\mathcal{A}_j\left(\omega - \widetilde{\omega}_{eg_j}\right)  = \sum_{i>j}\mathcal{A}_i J_{ij}+ \sum_{i<j}\mathcal{A}_i J_{ji} \nonumber\\
        &+ V_{bRj}\left(\frac{\tilde{t}_{bj} + \tilde{t}_{bj-1}}{2}\right)e^{ikx_j} + V_{bLj}\left(\frac{\tilde{r}_{bj} + \tilde{r}_{bj-1}}{2}\right)e^{-ikx_j}\nonumber \\ & + V_{tRj}\left(\frac{\tilde{t}_{tj} + \tilde{t}_{tj-1}}{2}\right)e^{imy_j} + V_{tLj}\left(\frac{\tilde{r}_{tj} + \tilde{r}_{tj-1}}{2}\right)e^{-imy_j}, \\
       & \tilde{t}_{bj}  = \tilde{t}_{bj-1} - i\frac{V_{bRj}}{v_{bg}} \mathcal{A}_je^{-ikx_j},\\
       & \tilde{r}_{bj+1}  = \tilde{r}_{bj} + i\frac{V_{bLj}}{v_{bg}} \mathcal{A}_j e^{ikx_j},\\
       & \tilde{t}_{tj}  = \tilde{t}_{tj-1} - i\frac{V_{tRj}}{v_{tg}} \mathcal{A}_je^{-imy_j},\\
       & \tilde{r}_{tj+1}  = \tilde{r}_{tj} + i\frac{V_{tLj}}{v_{tg}} \mathcal{A}_je^{imy_j}.
    \end{align}
\end{subequations}
This set of equations can be solved for any number of QEs (i.e., $j$ value with $1\leq j \leq \mathcal{N}$) coupled to our bidirectional waveguide ladder with the appropriate boundary conditions. The net transmission probability from the top and bottom waveguide end ports can be defined as $T_t=\left|\tilde{t}_{t\mathcal{N}}\right|^2$ and $T_b=\left|\tilde{t}_{b\mathcal{N}}\right|^2$. Similarly, the net reflection from Port 1 (bottom waveguide) and Port 3 (top waveguide) can be expressed as $R_t=\left|\tilde{r}_{t1}\right|^2$ and $R_b=\left|\tilde{r}_{b1}\right|^2$, respectively. In the next section, we present a numerical solution to the above set of transport equations for the QE chain consisting of up to twenty emitters. 

\begin{figure*}
\centering
\begin{tabular}{@{}cccc@{}}
\includegraphics[width=2.5in, height=1.65in]{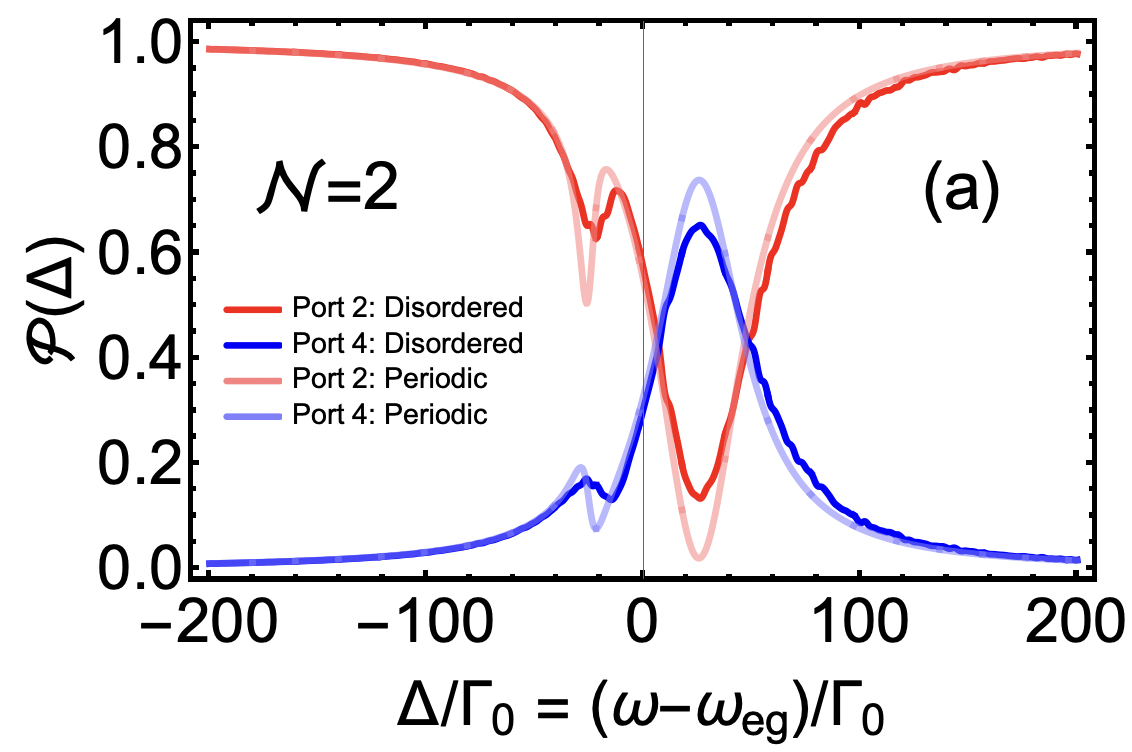} &
\hspace{-4.5mm}\includegraphics[width=2.4in, height=1.65in]{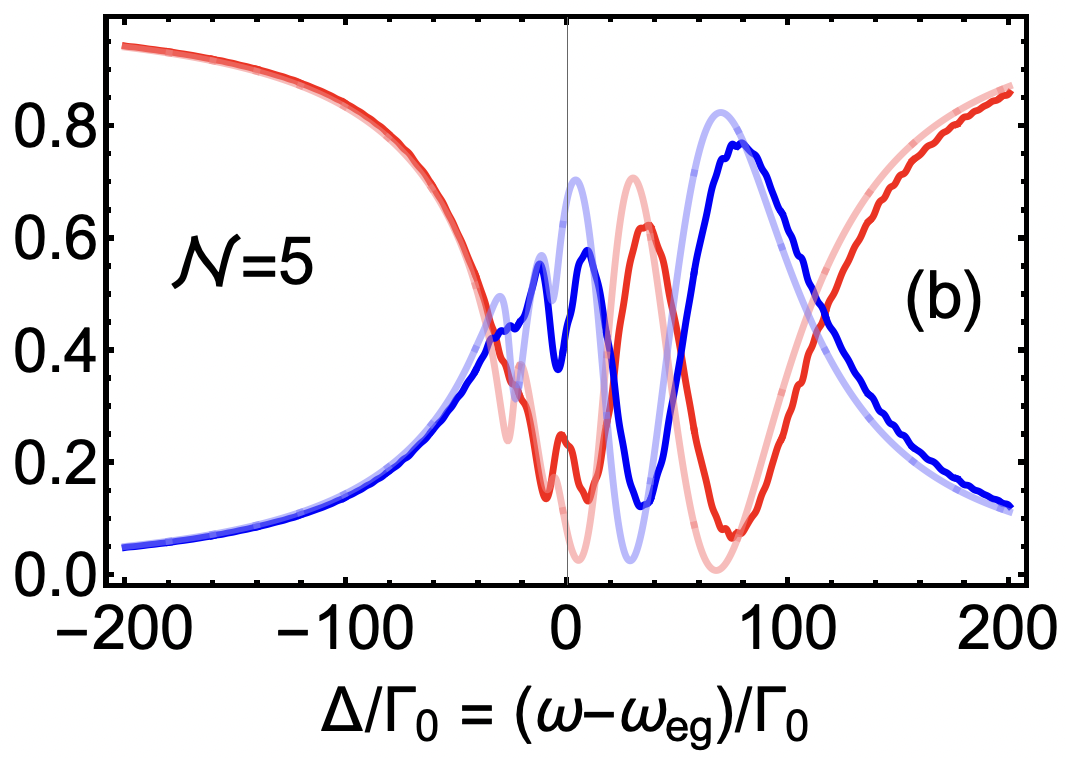} \\
\hspace{-1.5mm}\includegraphics[width=2.5in, height=1.65in]{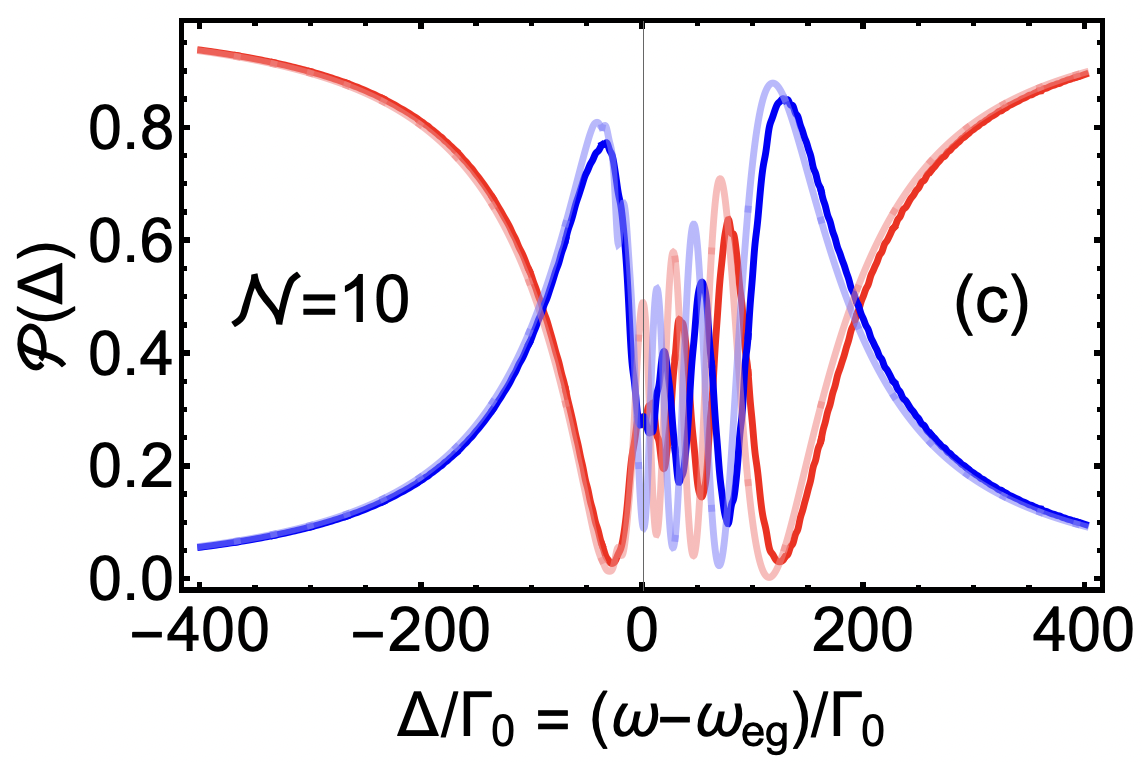} & \hspace{-1.5mm}\includegraphics[width=2.4in, height=1.65in]{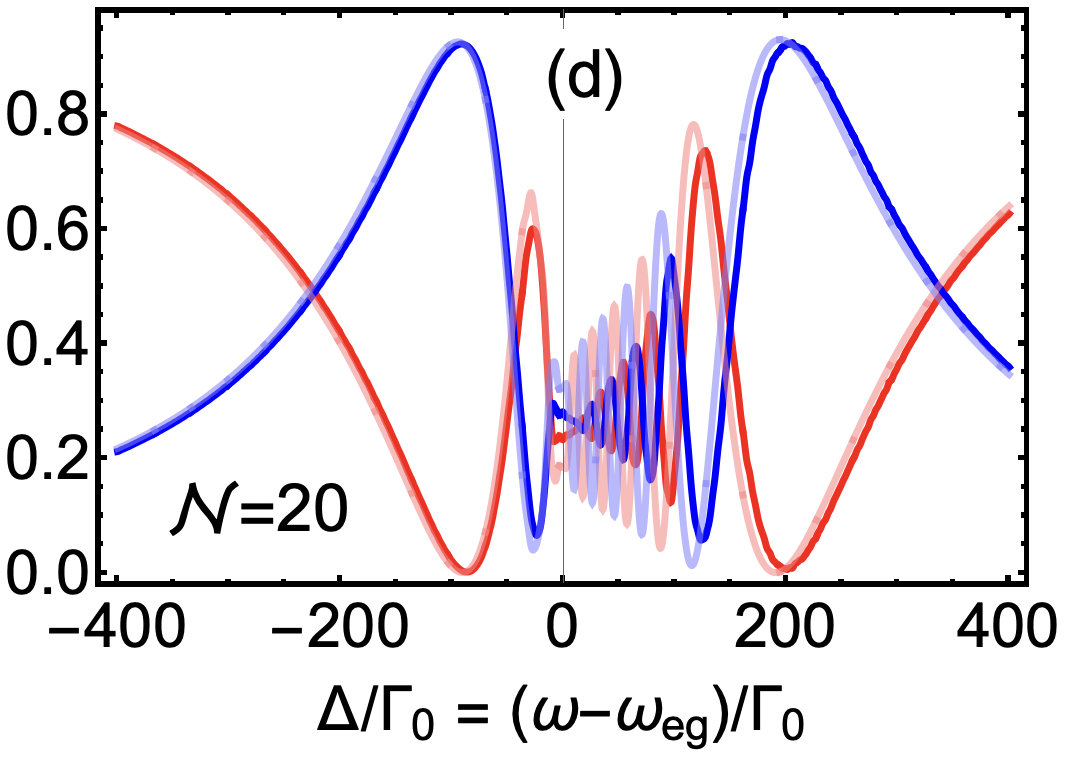}  
\end{tabular}
\captionsetup{
format=plain,
margin=1em,
justification=raggedright,
singlelinecheck=false
}
\caption{(Color online) The detection probability $\mathcal{P}$ at Port 2 with disordered emitter array (red solid curve), Port 4 with disordered emitter array (blue solid curve), Port 2 with periodic emitter array (red translucent curve), and Port 4 with periodic emitter array (blue translucent curve) as a function of the detuning $\Delta$ in units of free space decay rate of the emitters $\Gamma_0$. Total number of emitters in figure: (a) $\mathcal{N}=2$, (b) $\mathcal{N}=5$, (c) $\mathcal{N}=10$, and (d) $\mathcal{N}=20$. As mentioned at the beginning of the present section, the parameters used in these plots are $\Gamma = 11.03\Gamma_0$, $\gamma = 6.86 \Gamma_0$, $\lambda_e = 655{\rm nm}$, distance (or mean distance for the disordered case) between the emitters $\mu =\lambda_e /20$ with standard deviation of $\sigma = 0.1\mu$. Note that this value of lattice constant has been used in Eq.~\eqref{DDI} to obtain the value of DDI in these plots.}\label{Fig3}
\end{figure*}

\begin{figure*}
\centering
\begin{tabular}{@{}cccc@{}}
\includegraphics[width=2.5in, height=1.7in]{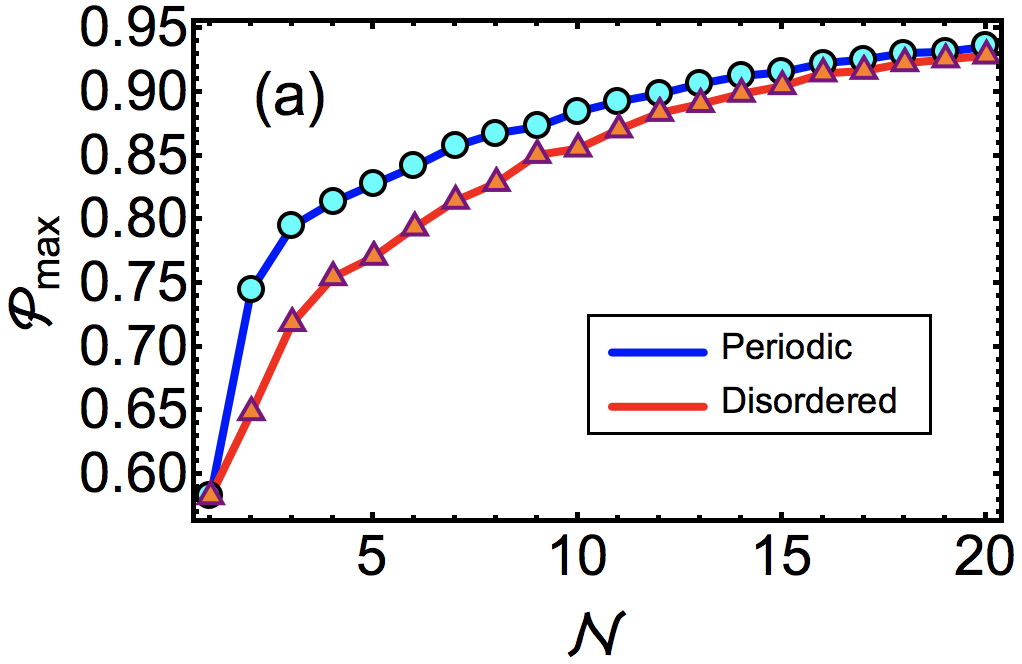} &
\hspace{-1.5mm}\includegraphics[width=2.5in, height=1.7in]{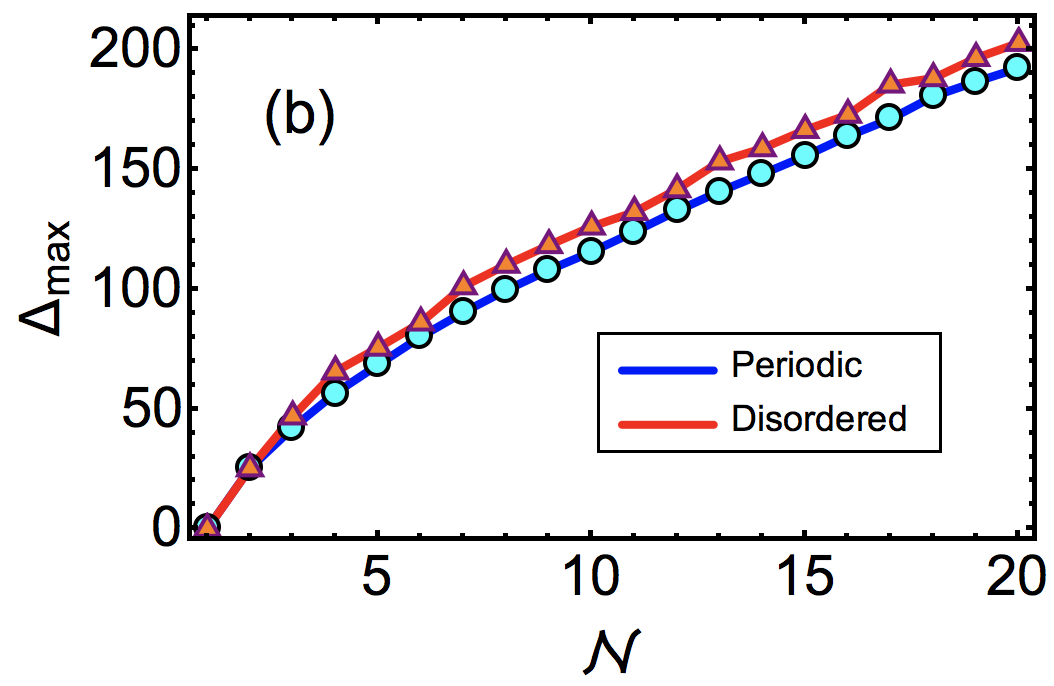} 
\end{tabular}
\captionsetup{
format=plain,
margin=1em,
justification=raggedright,
singlelinecheck=false
}
\caption{(Color online) (a) Maximum probability $\mathcal{P}_{\rm max}$ of detecting the single photon at Port 4 for periodic (blue curve with cyan dots) and disordered (red curve with orange triangles) case as a function of the number of QEs $\mathcal{N}$ in the wQED ladder. Note that, for each $\mathcal{N}$ case, the detuning $\Delta$ can be different at which $\mathcal{P}_{\rm max}$ is achieved. (b) The detuning value at which $\mathcal{P}_{\rm max}$ is found (we call it $\Delta_{\rm max}$ and it is measured in units of $\Gamma_0$) plotted against the emitter number $\mathcal{N}$. The rest of the parameters in both plots are the same as used in Fig~\ref{Fig3}.}\label{Fig4}
\end{figure*}

\section{\label{sec:level4} Results: Photon routing in many (upto twenty) emitter wQED ladders}
The theoretical model we have considered in this work can be applied to various experimental platforms suitable for wQED, as we have already pointed out in the first paragraph of the Introduction Section. To proceed with the results, we take the example of quantum dots coupled with nanowires and adopt the experimentally feasible parameters reported in Ref.~\cite{akimov2007generation, chang2006quantum}. In particular, we take emitter transition wavelength $\lambda_e  = 655{\rm nm}$, spontaneous emission rate $\gamma=6.86\Gamma_0$, the coupling strength between QEs and the waveguides $\Gamma=11.03\Gamma_0$, and DDI parameter $J=23.10\Gamma_0$. Note that the free space emitter decay rate $\Gamma_0$ has been taken as the unit for frequency in all results, which takes a value of $7.5{\rm MHz}$ in these experiments. Furthermore, even though the mathematical treatment presented in the last section works for bidirectional waveguide QED, building upon the work of Ref.~\cite{poudyal2020collective}, we now assume for the rest of the paper a strictly chiral regime in which all back reflections have been completely suppressed (as indicated in Fig.~\ref{Fig1} as well).

For both periodic and disordered cases, all QEs are assumed to be identical with equal coupling strength with both waveguides. The inter-emitter separation is $\lambda_e/20$, which avoids the involvement of any non-Markovian effects in both periodic and disordered cases \cite{zhang2014effects}. Finally, in the disordered case, the position disorder has been numerically generated by using a Gaussian probability distribution $\mathcal{G}$ of the form
\begin{align}
    \mathcal{G}(x) = \frac{1}{\sqrt{2\pi\sigma^2}}e^{-\frac{(x-\mu)^2}{2\sigma^2}},
\end{align}
with the mean inter-emitter distance $\mu$ and a standard deviation of $\sigma=0.1\mu$. Given the behavior of the disordered $J(L)$ curve observed in Fig.~\ref{Fig2}, we classify this as a weak disorder regime. 

\subsection{Transport probabilities}
We begin from a rather simple problem, namely, a single-photon routing in a wQED ladder with only two QEs. By setting the origin of coordinates at the location of the first QE, we call inter-emitter separation to be $r_{12}$, the transmission coefficient from Port 2 $\tilde{t}_{b2}$ and from Port 4 i.e. $\tilde{t}_{t2}$ takes the following form:
\begin{subequations}
    \begin{align}
    &\tilde{t}_{b_2} = \frac{4J^2+4\Gamma^2+\left(\gamma-2i\Delta\right)^2+8J\Gamma \sin\left(r_{12}\Delta\right)}{4J^2-8ie^{r_{12}\Delta}J\Gamma+\left(\gamma+2\Gamma-2i\Delta\right)^2},\\
    &\tilde{t}_{t_2} = \frac{4i\Gamma\left(i\gamma+2\Delta+2J\cos\left(r_{12}\Delta\right)\right)}{4J^2-8ie^{r_{12}\Delta}J\Gamma+\left(\gamma+2\Gamma-2i\Delta\right)^2},
\end{align}
\end{subequations}
where we have set $v_{b_g}=v_{t_g}=1$ such that $r_{12}\Delta$ which in fact is $r_{12}\Delta/v_g$ becomes dimensionless. In Fig.~\ref{Fig3}(a), we present the results for both periodic (translucent curves) and disordered (solid curves) configurations. In the $\gamma\neq 0$ and $J\neq 0$ scenario, the probability of detecting a single photon at Port 4 exhibits two asymmetric resonances around $\Delta=0$ point. As reported in Ref.~\cite{cheng2017waveguide}, such an asymmetry originates from energy decay due to atomic dissipation. From the routing perspective, we observe that in the periodic case, $\sim 74.6\%$ Port 1 to Port 4 routing has been achieved at $\Delta=25.3\Gamma_0$. On the other hand, the disorder suppresses the maximum value of Port 4 detection probability to $\sim 65.6\%$ at the same $\Delta$ value. 

\begin{figure*}
\centering
\begin{tabular}{@{}cccc@{}}
\includegraphics[width=2.45in, height=1.7in]{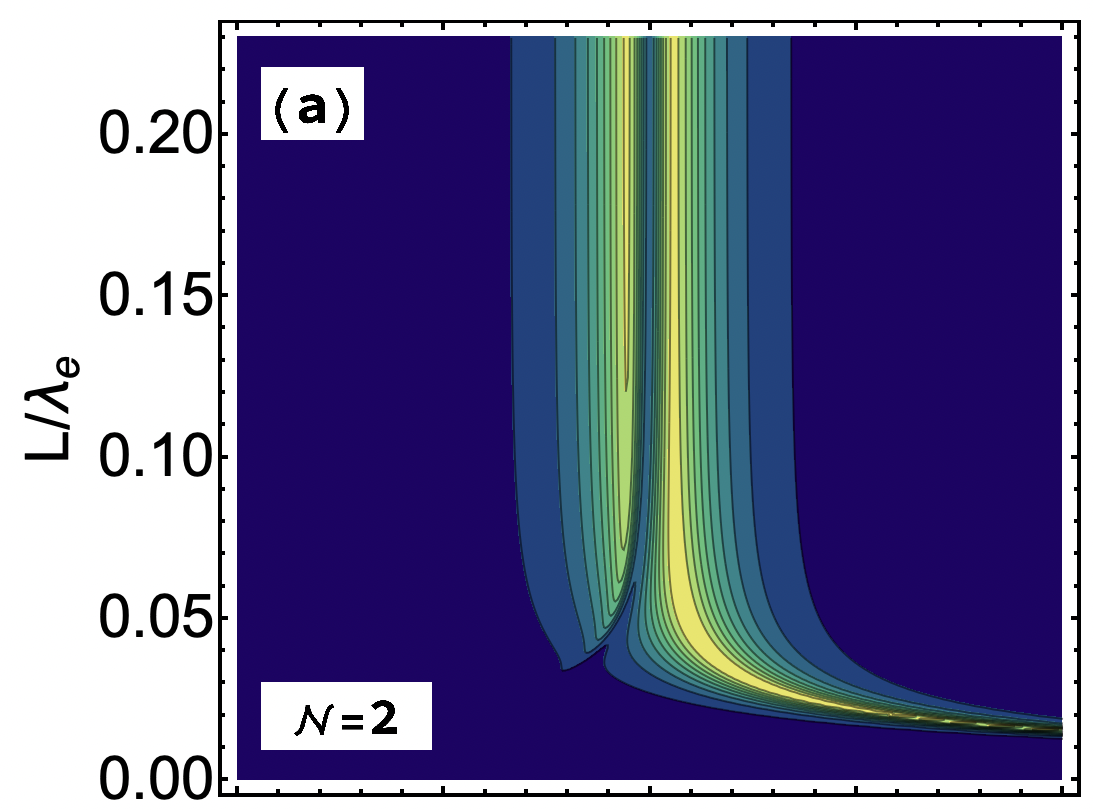} &
\includegraphics[width=2.4in, height=1.67in]{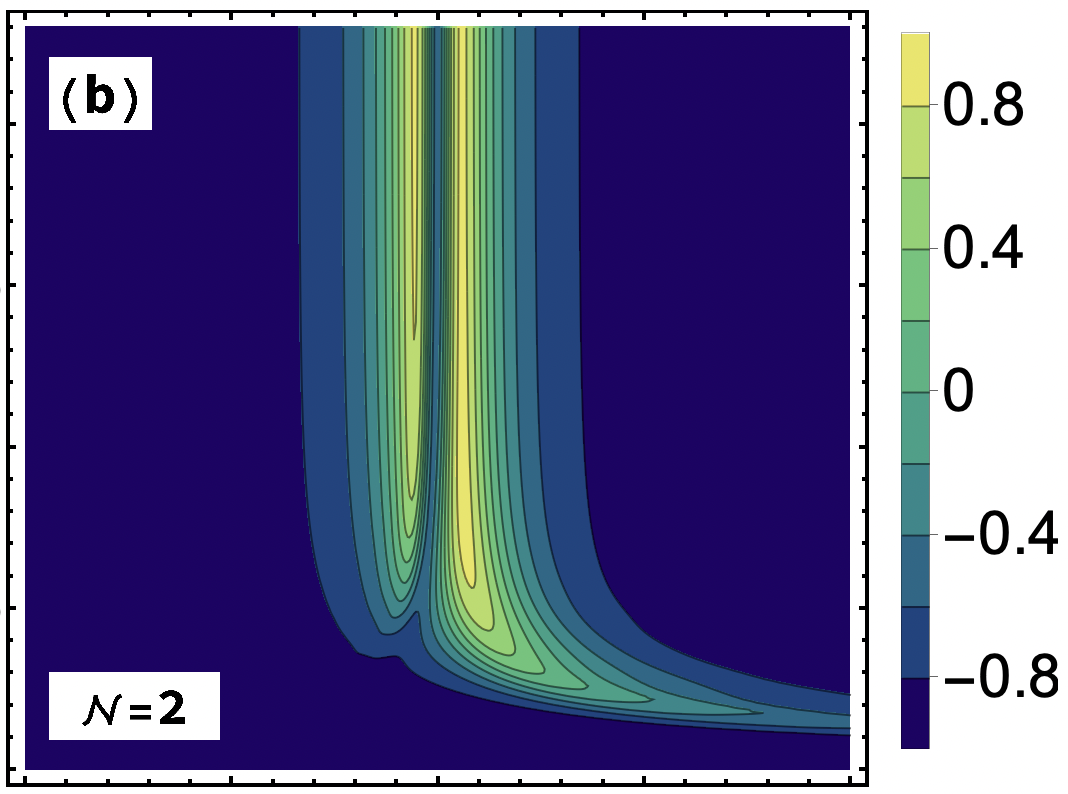} \\
\hspace{1mm}\includegraphics[width=2.5in, height=1.9in]{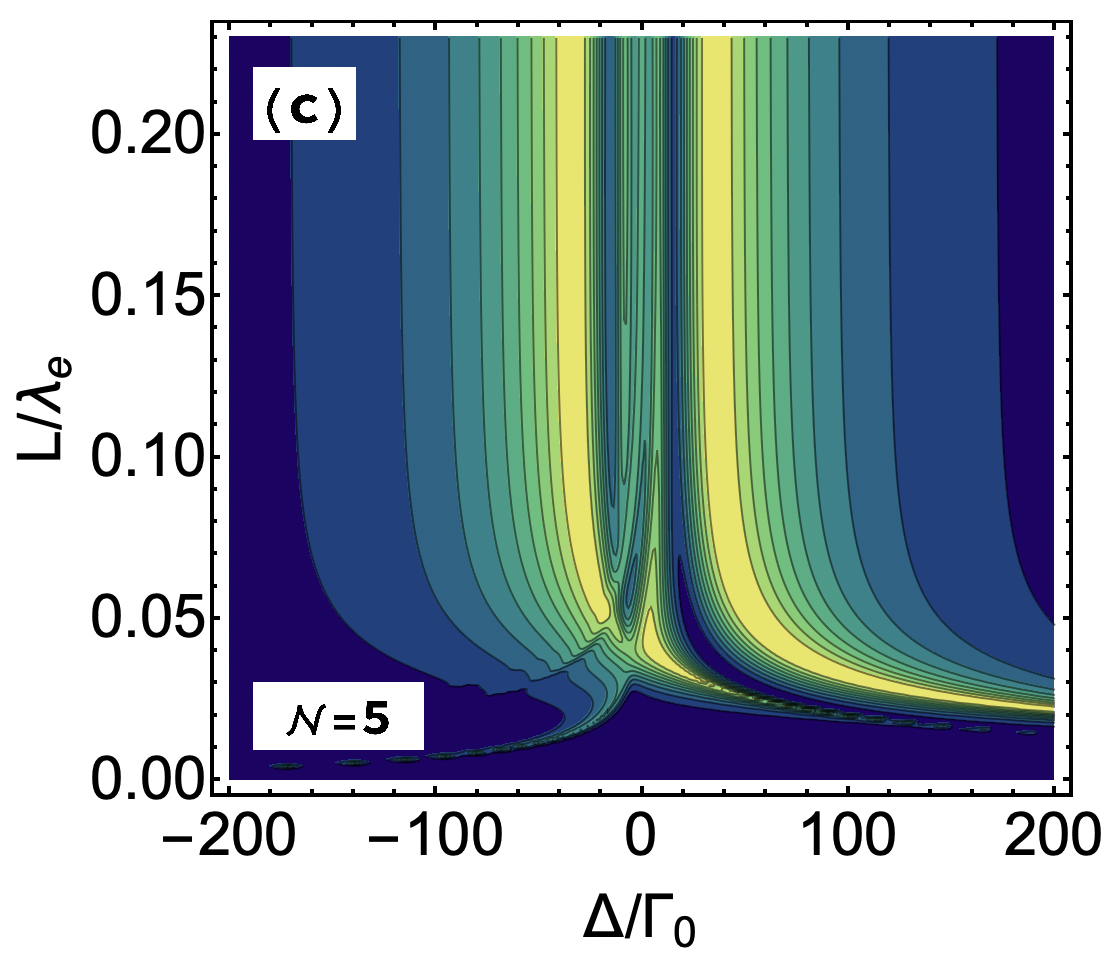} &
\hspace{-3mm}\includegraphics[width=2.5in, height=1.9in]{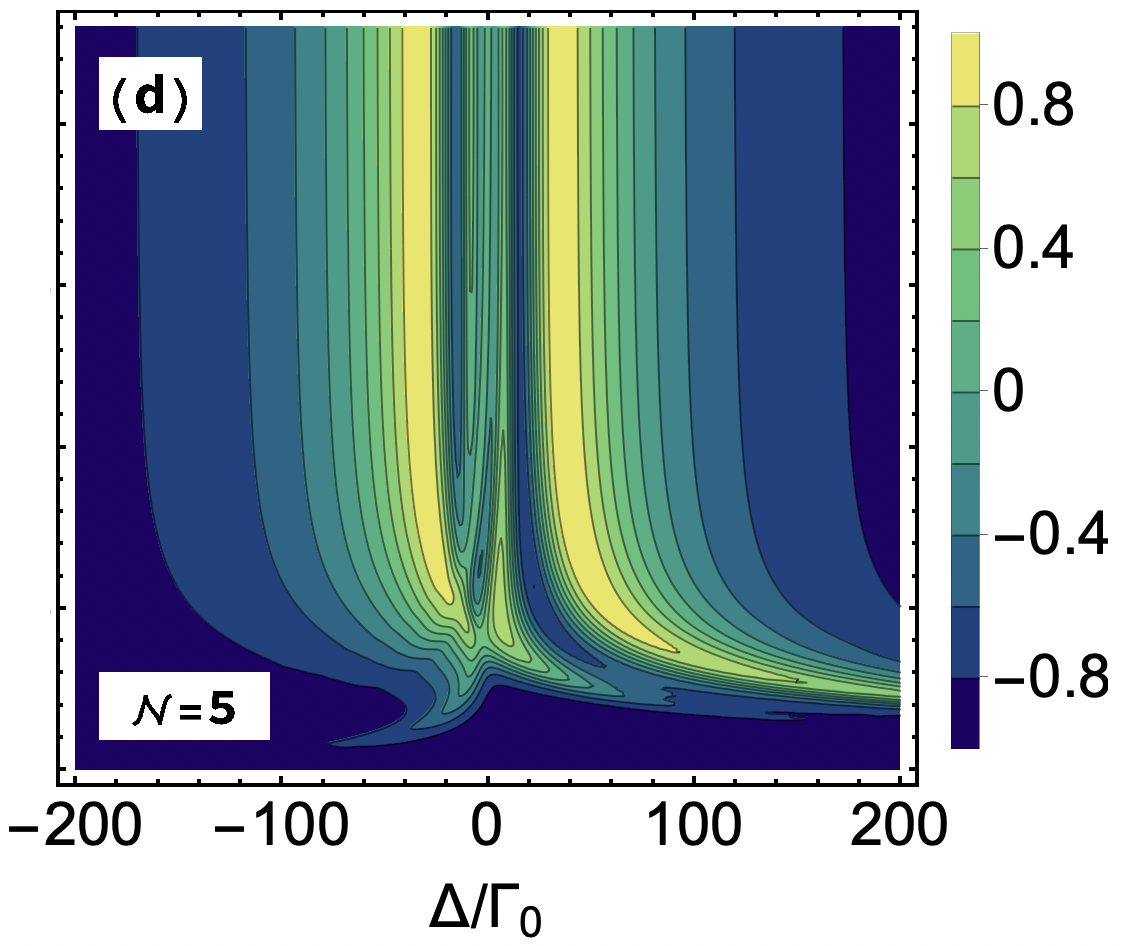}
\end{tabular}
\captionsetup{
format=plain,
margin=1em,
justification=raggedright,
singlelinecheck=false
}
\caption{(Color online) Density plots showing routing efficiency (through shades of blue, green, and yellow color shades) while varying detuning $\Delta$ and inter-emitter separation $L$. Plot (a) and (c) are $\mathcal{N}=2$ and $\mathcal{N}=5$ cases for periodic settings. At the same time, Plot (b) and (d) are $\mathcal{N}=2$ and $\mathcal{N}=5$ cases for the corresponding disorder configuration. The rest of the parameters are the same as assumed in Fig.~\ref{Fig3}(a) and Fig.~\ref{Fig3}(b).}\label{Fig5}
\end{figure*}

Next, in Fig.~\ref{Fig3}(b) to Fig.~\ref{Fig3}(d) we increase the number of QEs from $\mathcal{N}=5$ to $\mathcal{N}=20$ to promote collective emitter interactions. In both periodic and disordered settings, we find that with the increase in the QE number, the number of resonances increases in detection probabilities. As we approach $\mathcal{N}=20$, we notice the emergence of a pattern in detection probabilities where an asymmetric envelope of peaks is formed around $\Delta=0$ along with two major side peaks. These side peaks allow us to achieve a considerably higher degree of routing. For instance, for the periodic case, we find that for $\mathcal{N}=5$, $\mathcal{N}=10$, and $\mathcal{N}=20$ cases, the Port 4 probabilities respectively increase from $\sim 82.8\%$ to $\sim 88.4\%$ to $\sim 93.6\%$. We also note that these values of maximum probabilities exist at larger $\Delta$ values as $\mathcal{N}$ is increased.  

This trend of achieving higher Port 4 detection probabilities (improved routing) also extends to the disordered configuration. In the disordered case we observed for $\mathcal{N}=5$, $\mathcal{N}=10$, and $\mathcal{N}=20$ cases the corresponding Port 4 probabilities take the value of $\sim 77.8\%$ at $75.6\Gamma_0$, $\sim 85.5\%$ at $\Delta=126.23\Gamma_0$, and $\sim 92.9\%$ at $\Delta=202.9\Gamma_0$. Additionally, we make two important remarks. First, these higher Port 4 detection probabilities are accompanied by lower Port 2 probabilities, indicating genuine improvement in Port 1 to Port 4 routing as $\mathcal{N}$ increases. Secondly, as the QE number increases, the collective emitter effects shield the routing not only against environmental decoherence (spontaneous emission loss) but also against weak disorder in the emitter locations.

To further highlight the improvement in routing as a function of QE number, in Fig.~\ref{Fig4}(a), we plot the maximum detection probability at Port 4 (labeled as $\mathcal{P}_{\rm max}$) as a function of $\mathcal{N}$. The blue curve with cyan dots represents the periodic lattice case, while the red curve with orange triangles represents the corresponding disordered lattice. As expected, at $\mathcal{N}=1$, both periodic and disordered cases start from the same value of $\mathcal{P}_{\rm max}\sim 58\%$. However, as we increase the number of emitters, the maximum probability of redirection of single photons to Port 4 increases such that by the time we arrive at  $\mathcal{N}>15$, we find that $\mathcal{P}_{\rm max}$ reaches $90\%$. It is essential to notice that the periodic configurations always promote a higher value of $\mathcal{P}_{\rm max}$ as compared to the disordered case, but at $\mathcal{N}=20$, the difference between the periodic and disordered cases minimizes, thus, confirming the dominating nature of collective emitter effects against weak disorder. Next, in Fig.~\ref{Fig4}(b), we report the detuning values at which $\mathcal{P}_{\rm max}$ has been achieved. We notice that in periodic and disordered cases, the $\Delta_{\rm max}$ follows the (almost) same increasing trend as a function of $\mathcal{N}$.

\subsection{Routing efficiency}
\begin{figure*}
\centering
\begin{tabular}{@{}cccc@{}}
\includegraphics[width=3.5in, height=2.25in]{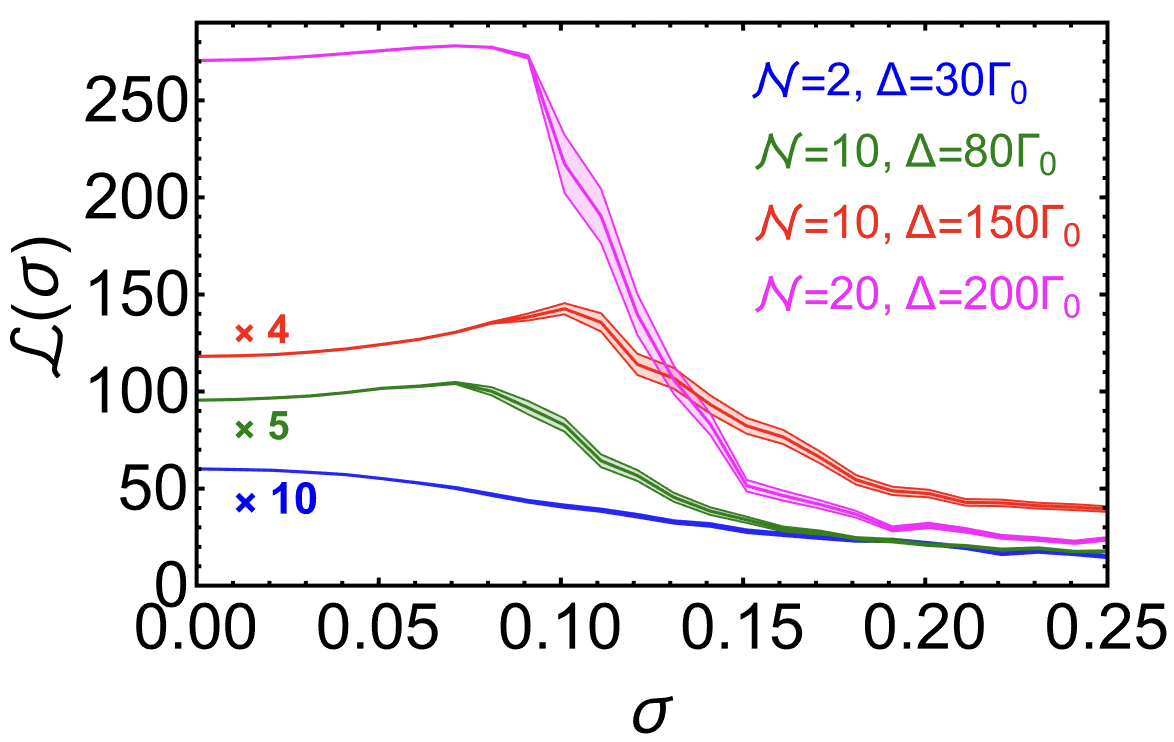}
\end{tabular}
\captionsetup{
format=plain,
margin=1em,
justification=raggedright,
singlelinecheck=false
}
\caption{(Color online) The localization length $\mathcal{L}$ approximated for a finite emitter chain (solid red curves) with errors (the surrounding lighter-colored regions) plotted against the disorder strength quantified through the standard deviation $\sigma$ in units of QEs mean position $\mu$. Four cases with $\mathcal{N} = 2$, $\Delta =30\Gamma_0$; $\mathcal{N} = 5$, $\Delta =80\Gamma_0$; $\mathcal{N} = 10$, $\Delta =150\Gamma_0$; and $\mathcal{N} = 10$, $\Delta =200\Gamma_0$ have been considered. Note that a different detuning value has been chosen in each curve, corresponding to the maximum routing efficiency. Some curves have been magnified to fit into the scale of the plot. The remaining parameters are the same as reported in Fig.~\ref{Fig3}}.\label{Fig6}
\end{figure*}
So far, in the previous results, we have considered a fixed inter-emitter separation of $\lambda_e/20$. However, in this work, it is crucial to analyze the impact of inter-emitter separation on the routing protocol. To this end, we introduce routing efficiency $\xi_\mathcal{N}$ which for a $\mathcal{N}$-emitter wQED ladder is defined as
\begin{align}
    \xi_\mathcal{N} = \frac{\left|\tilde{t}_{t\mathcal{N}}\right|^2 - \left|\tilde{t}_{b\mathcal{N}}\right|^2}{\left|\tilde{t}_{t\mathcal{N}}\right|^2+\left|\tilde{t}_{b\mathcal{N}}\right|^2},
\end{align}
where for no-loss problem $-1\leq \xi_\mathcal{N}\leq 1$ with $\xi_\mathcal{N}=-1$ indicates that the photon appears at Port 2 while $\xi_\mathcal{N}=+1$ means photon being routed to Port 4. When spontaneous emission loss is included, the higher positive value of $\xi_\mathcal{N}$ would indicate better routing. Explicitly, for $\mathcal{N}=2$ case, the routing efficiency takes the form
\begin{widetext}
\begin{align}
    \xi_2 = -\frac{4J^2+\gamma^2+4\gamma\Gamma+4\Gamma^2-4i\gamma\Delta-8i\Gamma\Delta-4\Delta^2-8iJ\Gamma \cos\left(r_{12}\Delta \right)+8 J\Gamma \sin\left(r_{12}\Delta \right)}{4J^2+\gamma^2-4\gamma\Gamma+4\Gamma^2-4i\gamma\Delta+8i\Gamma\Delta-4\Delta^2+8iJ\Gamma\cos\left(r_{12}\Delta\right)+8J\Gamma\sin\left(r_{12}\Delta\right)},
\end{align}
\end{widetext}
where the last equation holds for both periodic and disordered cases; in the periodic (disordered) case the parameters $r_{12}$ and $J$ are expected to remain fixed (varying).

In Fig.~\ref{Fig5}, we plot routing efficiency for a two- and a five-emitter ladder as a density plot while on the horizontal axis varying the detuning $\Delta$ from $-200\Gamma_0$ to $+200\Gamma_0$ and on the vertical axis the inter-emitter separation $L$ from $0.01\lambda_e$ to $0.23\lambda_e$. Both periodic and disordered situations have been presented side-by-side for both cases. We notice that in all four plots, for smaller values of $L$ (say around $L=0.025\lambda_e$) $\xi_{\mathcal{N}}$ exhibits branches of positive values of more than $0.8$ value on and on the right side of $\Delta=0$ point. Another similar feature is, as we increase $L$, the right branch of $\xi_{\mathcal{N}}>0$ tends to vanish, and we observe vertical bands of diminishing $\xi_{\mathcal{N}}$ values as we move away from $\Delta=0$. Finally, we find that the difference between periodic and disordered cases is most visible at smaller $L$ values (see, for instance, the behavior of $\xi_2$ between $0\leq \Delta\leq 100\Gamma_0$ at $L=0.025$). This trend is understandable, as the position disorder for wQED ladders with not a large number of QEs is expected to change the inter-emitter separation or DDI more, which in turn alters the transmission pattern from Port 4 -- an argument which is also supported by Fig.~\ref{Fig2}.

\subsection{Localization length}
To quantify the degree to which the single photon has been localized in our system due to disorder, we make use of the following definition of localization length $\mathcal{L}$ as reported to be used in one-dimensional electronic systems \cite{markos2008wave, delande2013many}
\begin{align}\label{loclen}
    \mathcal{L}^{-1} = \lim_{\mathcal{N}\rightarrow\infty}\frac{\langle T_t\rangle}{\mathcal{N}}.
\end{align}
Here, $T_t$ represents the transmission probability for Port 4. As evident from Eq.~\eqref{loclen}, the above definition becomes more and more exact as the number of QEs $\mathcal{N}$ becomes large. However, in the following, we'll consider the cases with QE numbers up to $\mathcal{N}=20$; therefore, the reported results would represent an approximate behavior of the localization length. 

In Fig.~\ref{Fig6}, we plot the localization length $\mathcal{L}$ as a function of disorder strength (quantified through the standard deviation $\sigma$) for emitter chains with $\mathcal{N} = 2, 5, 10,$ and $20$ QEs. We notice that for relatively weaker disorder (when $\sigma\lesssim 0.075\mu$), the localization length takes a larger value in every case. This indicates that longer emitter chains can localize the photon propagation for the lower disorder. However, as we consider $\sigma>0.1\mu$, irrespective of $\mathcal{N}$, $\mathcal{L}$ decays and reaches considerably lower values (see the behavior of all curves around $\sigma\sim 0.2\mu$), this behavior contrasts with Anderson localization observed in traditional electronic systems with the disorder (for example, semiconductor materials with defects) where strong disorder promotes halting of electron transport \cite{wiersma1997localization}. Since the DDI has a sensitive dependence on the inter-emitter separation in our case, we argue that with the increasing disorder in our wQED ladder setup, the average inter-emitter separation can be decreased (in some iterations of the averaging process), which will have a drastic effect on DDI. In such cases, we expect the collective emitter effects to be enhanced and lead to improved photon transport (or less photon localization) through the DDI channel, thus producing this non-intuitive trend.

\section{\label{sec:level5} Summary and Conclusions}
This paper has studied the interplay between collective emitter effects and position disorder on single photon routing in many (up to $N=20$) emitter chiral wQED ladders. Using the machinery of the real-space formalism of quantum optics combined with Bethe ansatz, we derived single-photon transport equations for any number of QEs. We numerically solved those equations and plotted the results for ladders with $\mathcal{N}=2$, $\mathcal{N}=5$, $\mathcal{N}=10$ and $\mathcal{N}=20$ DDI QE chains. Results have been reported for periodic and position-disordered wQED ladder lattices (subjected to weak disorders) for quantum dots coupled to ${\rm Ag}$ nanowire experimental platform. The main conclusions of this work are summarized as follows.

From the transport probabilities results, in both periodic and disordered cases, we find a considerable improvement in the maximum value of Port 4 detection probability from 58\% to more than 90\% as the QE number is increased from one to twenty. As we compare the periodic versus disordered cases, we find that for $\mathcal{N} < 15$, the periodic case promotes a higher value of Port 4 maximum detection probabilities. However, as we reach $\mathcal{N}=20$, the difference between periodic and disordered case Port 4 maximum probability tends to get considerably reduced. This numerical finding indicates that the collective emitter effects dominate at that level and protect the routing scheme against environmental losses and weak position disorders in the wQED ladders.

Finally, the localization behavior as a function of disorder strength revealed a considerable decrease in the localization length for relatively high disorders (for instance, when $\sigma=0.15\mu$) for all cases of $\mathcal{N}$. We attributed this trend (which contradicts the standard Anderson localization in electronic systems) to the dominating nature of collective DDI effects compared to destructive interference in probability amplitudes due to position disorder.

\section*{Acknowledgements}
This work is supported by the NSF Grant \# LEAPS-MPS 2212860 and the Miami University College of Arts and Science \& Physics Department start-up funding.

\bibliographystyle{ieeetr}
\bibliography{paper}
\end{document}